\documentclass[seAMS,usenatbib]{mn2e}
\usepackage{graphicx}
\usepackage{epstopdf}
\usepackage{amsmath}

\usepackage[hyphens]{url}
\usepackage{color}
\usepackage[breaklinks,colorlinks=true,citecolor=cobalt,linkcolor=alizarin,urlcolor=cobalt]{hyperref}
\usepackage[hyphenbreaks]{breakurl}
\definecolor{cobalt}{rgb}{0.0, 0.28, 0.67}
\definecolor{airforceblue}{rgb}{0.36, 0.54, 0.66}
\definecolor{armygreen}{rgb}{0.29, 0.33, 0.13}
\definecolor{alizarin}{rgb}{0.82, 0.1, 0.26}

\usepackage[labelformat=empty,labelsep=none]{subcaption}
\newcommand{\dummytitle}[1]{}

\newcommand{\bhn}{10^5\, h^{-1}\, \rm{M}_{\sun}}
\newcommand{\dmn}{10^{10}\, h^{-1}\, \rm{M}_{\sun}}
\newcommand{\sm}{10^3\, h^{-1}\, \rm{M}_{\sun}}
\newcommand{\mbh}{M_{\rm{BH}}}
\newcommand{\mdm}{M_{\rm{DM}}}
\newcommand{\mbhdm}{M_{\rm{BH}}^{\rm{DM}10}}
\newcommand{\mgasdm}{M_{\rm{gas}}^{\rm{DM}10}}
\newcommand{\mdmbh}{M_{\rm{DM}}^{\rm{BH}5}}
\newcommand{\zdm}{z^{\rm{DM}10}}
\newcommand{\zbh}{z^{\rm{BH}5}}

\newcommand{\mgdm}{M_{\rm{gas}}^{\rm{DM}10}}

\newcommand{\dg}{\rho_g > \rho_c}
\newcommand{\msun}{\rm{M}_{\sun}}

\begin{document}
\title[Black hole seeding in cosmological simulations]{The impact of black hole seeding in cosmological simulations}
\author[E.~Wang, P.~Taylor, C.~Federrath, and C.~Kobayashi]{Ella~Xi~Wang$^1$\thanks{Email: ellawang@mso.anu.edu.au}, Philip~Taylor$^{1, 2}$, Christoph~Federrath$^{1}$, Chiaki~Kobayashi$^3$ \\
$^1$Research School of Astronomy and Astrophysics, Australian National University, Canberra, ACT 2611, Australia\\
$^2$ARC Centre of Excellence for All Sky Astrophysics in 3 Dimensions (ASTRO 3D), Australia\\
$^3$Centre for Astrophysics Research, School of Physics, Astronomy and Mathematics, University of Hertfordshire, Hatfield, AL10 9AB, UK}
\maketitle

\begin{abstract}
Most cosmological simulations of galaxy evolution include active galactic nucleus (AGN) feedback, typically seeding black holes with masses of $\geq \bhn$ when the dark matter halo exceeds a given threshold mass. \citet{taylor14} introduced a new model, which seeds black holes at $\sm$ based on gas properties alone, and motivated by the channel of black hole formation due to the collapse of the most massive first stars in the Universe. We compare the black hole mass when the dark matter halo mass is $\dmn$ between the different seeding methods. We find that seeding based upon gas properties gives a distribution of black hole masses with $\langle \log \mbh {/ \msun} \rangle = (5.18 \pm 0.54)$ when dark matter halo mass is $\dmn$, consistent with the {seeding criteria} used in other simulations. However, the evolution of individual galaxies can be strongly affected by the different seeding mechanisms. We also find that the mean value of the distribution of black hole masses at a given halo mass evolves over time, with higher masses at higher redshifts, indicative of downsizing. Our results can inform more physically motivated black hole and AGN feedback models in cosmological simulations and semi-analytic models.
\end{abstract}


\section{Introduction}
The $\Lambda$ Cold Dark Matter ($\Lambda$CDM) model is the currently accepted model of the structure we see in the present-day Universe. Cosmological simulations are based on the $\Lambda$CDM model as it successfully accounts for large-scale structure and the distribution of galaxies in the universe, the primordial abundance of hydrogen and helium, and the accelerating expansion of the Universe.

Active galactic nuclei (AGN) feedback is required in cosmological simulations to reproduce certain observations: cosmic star formation rate (SFR) \citep{booth09, vogelsberger14, taylor14}, various downsizing phenomena - including the [$\alpha$/Fe] relation of early type galaxies and mass and redshift dependences of the specific SFRs \citep{juneau05, thomas05, stark13}. $\Lambda$CDM predicts that galaxies grow hierarchically: smaller galaxies are formed in dark matter haloes and merge to form bigger galaxies. This implies that the most massive galaxies should be younger and have high SFR. However, observations show that they, in fact, have low SFR at the present day and older stellar populations. This is known as downsizing, and may be influenced by AGN feedback. {In hydrodynamical simulations, AGN feedback can effectively quench star formation in massive galaxies.} Thus, these large galaxies contain predominantly older stars once star formation is quenched. Mergers can suppress star formation by triggering AGN activity \citep[e.g.,][]{hopkins08}, lowering the SFR of massive galaxies.

Galaxy evolution can be studied via observations or simulations. Observations show us galaxy properties at one snapshot in time, for a large number of galaxies. However, we can only see individual galaxies at a single point in their evolution. Simulations give us the full history of a galaxy, but are limited by the number of galaxies that can be simulated. Semi-analytical models apply prescriptions for baryonic physics to pre-computed dark matter-only haloes and merger trees
\citep{kauffmann93, cole94, bower06, croton16}. This type of model has the advantage of being able to run large suites of models in relatively little time. The other method is to simulate baryonic processes directly, together with the structure formation by dark matter. This is more time-consuming, but the baryons and dark matter haloes evolve self-consistently.

The mass of a black hole helps determine its accretion rate, and, in turn, its feedback energy, which means that black holes co-evolve with their host galaxy. Many correlations between properties of the host galaxy and the central black hole are observed{: bulge stellar mass} \citep{magorrian98, ferrarese00, gebhardt00, tremaine02, marconi03, haring04, sani11}, stellar velocity dispersion \citep{ferrarese00, gebhardt00, tremaine02, kormendy13}, luminosity \citep{kormendy95, marconi03}, and S\'ersic index \citep{graham01, graham07, savorgnan13}. Large volume hydrodynamical simulations (e.g., Illustris \citep{vogelsberger14}, IllustrisTNG \citep{weinberger17}, and EAGLE \citep{schaye15}) place black hole seeds in massive dark matter haloes, then grow the black hole and introduce feedback. \citet{taylor14} adopted a different approach by seeding black holes based on gas properties {(see Section 2 for more details)}, similar to the approach taken for Horizon-AGN \citep{dubois12, dubois14} {and SuperChunky \citep{habouzit17}.} Both black hole seeding models can successfully reproduce observed properties {including those that depend directly on black hole mass, such as the Magorrian relation \citep{magorrian98}, and indirectly, such as the stellar mass and star formation rate. However, black holes are seeded at different times in these models, and it is not clear how the different accretion and feedback histories can affect the evolution of individual galaxies. In this paper, we compare the black hole masses in dark matter haloes from our simulations to the black hole mass that would have been seeded by the method based on dark matter halo properties. In order to compare the two seeding methods fairly, we run a simulation with the same initial conditions and identical physics; apart from the seeding prescription. This method of comparison removes the complex effects of the different physical models of stellar and AGN feedback used in our simulations and others.}

In this paper, we compare the seeding method used in the \citet{taylor14} model against those of EAGLE, Illustris, and IllustrisTNG. In Section~\ref{sim}, we introduce the black hole seeding models in more detail, along with the comparison method. In Section~\ref{res}, we present the results. Lastly, in Section~\ref{diss}, we discuss the implications of these results, and investigate for the spread of masses. The results of these comparisons can inform future models of black hole seeding.

\section{Methods}
\label{sim}
\subsection{Our simulation}
\begin{figure}
	\centering
	\includegraphics[width=0.48\textwidth,keepaspectratio]{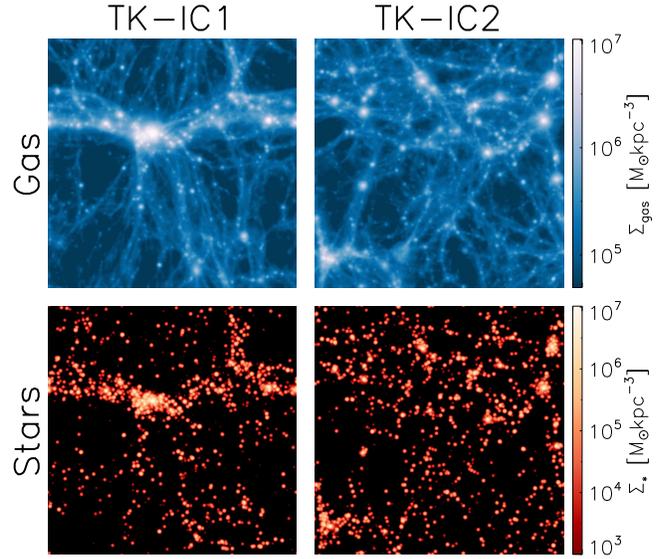}
	\caption{{Maps of surface density of gas (top row) and stars (bottom row) for simulations TK-IC1 (left column) and TK-IC2 (right column).}}
	\label{fig:TK_IC_comp}
\end{figure}

The simulation presented in \citet{taylor15a}, which we shall subsequently refer to as TK-IC1, is based on the smoothed particle hydrodynamics (SPH) code \textsc{gadget-3} \citep{springel05}. The code has fully adaptive individual smoothing lengths and timesteps, and uses the entropy conserving formulation of SPH \citep{springel02}. Included in the simulation are baryonic processes relevant for galaxy formation and evolution: radiative cooling \citep{sutherland93}, star formation \citep{ck04, ck07}, chemical enrichment \citep{ck06, ck09, ck11b}, supernova feedback \citep{ck07}, and black hole physics \citep{taylor14}. {We employ a 9-year \emph{Wilkinson Microwave Anisotropy Probe} $\Lambda$CDM cosmology \citep{hinshaw13} with $h=0.70$, $\Omega_{\rm{m}} = 0.28$, $\Omega_{\Lambda}=0.72$, $\Omega_{\rm{b}}=0.046$, and $\sigma_8=0.82$. The simulation box is 25 $h^{-1}$ Mpc (comoving) on a side with $240^3$ particles of each of dark matter and gas, with masses $M_{\rm{DM}} = 7.3 \times 10^7 \, \msun$ and $M_{\rm{gas}}=1.4 \times 10^{7} \, \msun$. Photo-heating is given by a uniform and time-evolving UV background radiation field \citep{haardt96}. We also analyse a simulation with different initial conditions but identical physics; this simulation is referred to as TK-IC2, which represents a less clustered environment. Fig.~\ref{fig:TK_IC_comp} shows the gas and stellar surface density of TK-IC1 (left panels) and TK-IC2 (right panels).}

\subsection{Black hole seeding in TK}
Due to the theory of primordial star formation and the observed first chemical enrichment, the most likely candidates for seed black holes {\citep{volonteri10, latif16}} are the remnants of Population \textsc{iii} stars {\citep{madau01, bromm02, schneider02}}, the direct collapse of primordial gas {\citep{bromm03, koushiappas04, agarwal12}, or via a massive quasi-star formed from collisions in nuclear star clusters in the early Universe \citep{baumgarte99, portegies99, devecchi09}}. In all black hole formation mechanisms, dense gas and inefficient cooling (to prevent {fragmentation down to solar mass stars}) are required. Thus, in \citet{taylor14}, black hole formation criteria used were such that black holes can form directly based on local gas properties, if gas particles satisfy:
\begin{equation}
\dg \textrm{ and } Z=0,
\end{equation}
where $\rho_g$ is the gas particle density, $\rho_c$ is a specified critical density, and $Z$ is the gas metallicity. {The gas density is calculated using the SPH kernel, chemical composition information is recorded on a particle-by-particle basis.} In the data set analysed, $\rho_{\rm c} = 0.1\, h^2\, m_{\rm H}\, {\rm cm}^{-3}$ \citep{taylor14}. Once these criteria are satisfied, the gas particle is converted into a black hole with seed mass $\sm$. Seed black holes grow via gas accretion and mergers. {There is no restriction on the number of black holes formed per halo.}

\subsection{Black hole seeding in other cosmological simulation suites}
{Some other cosmological simulations, such as Horizon-AGN and SuperChunky, take a similar approach to TK and seed black holes based on gas criterion, whilst others such as EAGLE and Illustris seed black holes once the dark matter halo reaches a mass threshold. {Horizon-AGN seeds black holes in dense regions at the center of galaxies using a seed mass of $10^5 \, \msun$; a maximum of one black hole is seeded per galaxy \citep{dubois12, dubois14}.} SuperChunky seeds black holes in dense, low metallicity regions ($Z<10^{-3.5} \, {\rm Z}_{\odot}$), and uses a seed mass based upon an initial mass function \citep{habouzit17}. These simulations differ only in their details from TK-IC1; the focus of this paper is to compare to simulations that use halo properties to seed black holes.}

We compare our black hole and dark matter masses against Illustris, IllustrisTNG, and EAGLE. Illustris seeds black holes when $\mdm = 5 \times \dmn$ (where $\mdm$ is the dark matter mass), with a seed mass of $\mbh = \bhn$ \citep{sijacki15}. IllustrisTNG seeds black holes when $\mdm = 5 \times \dmn$, with seed mass $\mbh = 8 \times \bhn$ \citep{weinberger17}. EAGLE seeds black holes when $\mdm = \dmn$, with a seed mass of $\mbh = \bhn$ \citep{schaye15}. A summary of the different black hole seeding criteria and seed masses in the different simulation suites is provided in Table~\ref{tab:seeding}.

{To seed based upon dark matter halo mass, dark matter haloes need to be identified on the fly. In EAGLE, Illustris, and IllustrisTNG,} dark matter haloes are identified with a Friends-of-Friends (FoF) algorithm; typically dark matter particles within 0.2 times the mean separation of all dark matter particles are grouped together \citep{schaye15}. If a halo exceeds a given threshold mass and does not already contain a black hole, a gas particle within the halo is converted to a black hole with the specified seed mass.

\begin{table*}
\setlength{\tabcolsep}{5.0pt}
\begin{tabular}{lccccc}
\hline
& TK-IC1 & TK-IC2 & Illustris & IllustrisTNG & EAGLE \\
\hline
Seed mass & $\sm$ & $\sm$ & $\bhn$ & $8 \times \bhn$ & $\bhn$ \\
Seeding condition & $\dg$, $Z = 0$ & $\dg$, $Z = 0$ & $\mdm = 5 \times \dmn$ & $\mdm = 5 \times \dmn$ & $\mdm = \dmn$ \\
\hline
\end{tabular}
\caption{Black hole seed masses and seeding conditions for the different simulations compared in this paper. $\dg$ and $Z = 0$ are the seeding conditions for TK-IC1 and TK-IC2; while Illustris, IllustrisTNG, and EAGLE seed black holes based on a dark matter halo mass threshold.}
\label{tab:seeding}
\end{table*}

\subsection{AGN feedback in TK}
{In TK-IC1, the gravitational softening length of the simulation is $1.125\, h^{-1}\, \rm{kpc}$, so the small-scale physics of accretion onto black holes is not resolved; therefore,} we model the AGN feedback using the Eddington-limited Bondi-Holy accretion rate, given by:
\begin{equation}
\dot{M}_{\rm{acc}} = \textrm{min}(\dot{M}_{\rm{Bondi}}, \dot{M}_{\rm{Edd}}).
\label{accretion_rate}
\end{equation}
Bondi-Hoyle and Eddington accretion rates are given by:
\begin{equation}
\dot{M}_{\rm{Bondi}} = \alpha \frac{4\pi G^2M_{\rm{BH}}^2 \rho}{(c_{\rm{s}}^2 + v^2)^{3/2}}
\label{bondi}
\end{equation}
\begin{equation}
\dot{M}_{\rm{Edd}} = \frac{4\pi GM_{\rm{BH}}m_{\rm{p}}}{\epsilon_{\rm{r}} \sigma_{\rm{T}} c}
\label{eddington}
\end{equation}
where $\mbh$ is the black hole mass, $G$ is the gravitational constant, $c$ is the speed of light in a vacuum, $\rho$ is the gas density local to the black hole, $\sigma_{\rm{T}}$ is the Thompson cross section, $c_{\rm{s}}$ is the sound speed of the gas local to the black hole, $v$ is the relative velocity between the black hole and local gas, $m_{\rm{p}}$ is the proton mass, and $\alpha$ is a factor due to finite resolution of simulations. In TK-IC1, $\alpha = 1$.

Accreted material radiates energy as it falls onto a black hole, leading to a self-regulated accretion rate \citep{taylor14} and reduced SFR in massive galaxies \citep{taylor15a, taylor16, taylor17}. In each timestep $\Delta t$, a black hole produces an amount of feedback energy $E_{\rm{FB}}$, which is calculated using:
\begin{equation}
E_{\rm{FB}} = \epsilon_{\rm{r}} \epsilon_{\rm{f}} \dot{M}_{\rm{acc}} c^2 \Delta t,
\label{feedback}
\end{equation}
where $\epsilon_{\rm{f}}$ is the fraction of radiated energy that couples to the gas, and $\epsilon_{\rm{r}}$ is the radiative efficiency of the black hole. For this simulation, $\epsilon_{\rm{f}} = 0.25$ \citep{taylor14}, and $\epsilon_{\rm{r}} = 0.1$ \citep{shakura73} are adopted. {The feedback energy is purely thermal and isotropic, distributed kernel-weighted to gas neighbour particles; the number of feedback neighbours is $N_{\rm{FB}} = 72$, which is the same for supernova feedback. The time-dependence of the feedback follows directly from the time-dependence of the accretion rate.

\subsection{AGN feedback in other cosmological simulation suites}
All these simulations use Eddington-limited Bondi-Hoyle accretion to model the black hole accretion rate \citep{springel05acc} as we do here (equations \eqref{accretion_rate} - \eqref{eddington}). The feedback energy in these simulations follows the same functional form as equation~\eqref{feedback}, with $E_{\rm{FB}} \propto \dot{M}_{\rm{acc}}$. Other simulation suites use a modified Bondi-Hoyle accretion rate, different constants in the feedback energy, and different feedback modes. Different simulations use different $\alpha$ values in the Bondi-Hoyle accretion rate (equation~\eqref{bondi}) to account for the finite resolution of simulations. TK and IllustrisTNG both do not modify the Bondi-Hoyle equation and use $\alpha = 1$ \citep{taylor14, weinberger17}; whilst Illustris uses $\alpha = 100$ \citep{sijacki15}, and EAGLE uses $\alpha = \min(C^{-1}_{\rm{visc}}(c_s/V_{\phi})^3$, which is a non-constant factor equivalent to the ratio of the Bondi and viscous time scales \citep{schaye15}. The constants $\epsilon_{\rm{r}}$ and $\epsilon_{\rm{f}}$ in the feedback energy (equation~\eqref{feedback}) also vary between simulations. TK and EAGLE use $\epsilon_{\rm{r}} = 0.1$, whilst Illustris and IllustrisTNG use $\epsilon_{\rm{r}} = 0.2$. The way feedback energy is injected into the surrounding ISM also differs, this is determined by the feedback mode. TK and EAGLE only implement a thermal feedback mode, where $\epsilon_{\rm f, therm} = 0.25$ in TK and $\epsilon_{\rm f, therm} = 0.15$ in EAGLE; whilst Illustris and IllustrisTNG implement both thermal and kinetic feedback modes; with $\epsilon_{\rm f, therm} = 0.05$, $\epsilon_{\rm{f,kin}} = 0.35$ in Illustris; and $\epsilon_{\rm f, therm} = 0.35$, $\epsilon_{\rm{f,kin}} = 0.2$ in IllustrisTNG. These differences are summarised in Table~\ref{tab:agn}.

\begin{table*}
\setlength{\tabcolsep}{5.0pt}
\begin{tabular}{c | c | c | c | c | c | c}
\hline
& Feedback & Effective $\alpha$ & $\epsilon_{\rm r}$ & $\epsilon_{\rm f, therm}$ & $\epsilon_{\rm{f,kin}}$ \\
\hline
TK & Thermal & 1 & 0.1 & 0.25 & - \\
Illustris & Thermal, kinetic, and radiative & $100$ & 0.2 & 0.05 & 0.35 \\
IllustrisTNG & Thermal and kinetic & $1$ & 0.2 & 0.1 & 0.2 \\
EAGLE & Thermal & $\min(C^{-1}_{\rm{visc}}(c_s/V_{\phi})^3, 1)$ & 0.1 & 0.15 & - \\
\hline
\end{tabular}
\caption{{Modified Bondi-Hoyle, feedback energy constants, and feedback modes for TK, Illustris, IllustrisTNG, and EAGLE - the simulations compared in this paper. Effective $\alpha$ refers to the $\alpha$ term in the Bondi-Hoyle equation (equation~\eqref{bondi}). $\epsilon_{\rm{r}}$ is the radiative efficiency of the black hole. $\epsilon_{\rm{f,therm}}$ is the fraction of energy emitted thermally. $\epsilon_{\rm{f,kin}}$ is the fraction of energy emitted kinetically. The factor $C^{-1}_{\rm{visc}}(c_s/V_{\phi})^3$ is equivalent to the ratio of the Bondi and viscous time scales \citep{schaye15}.}}
\label{tab:agn}
\end{table*}

From equation \eqref{feedback} it can be seen that the product $\epsilon_{\rm r}\epsilon_{\rm f}$ determines the amount of energy that couples to the gas.
This value differs by at most a factor of 2.5 between the simulations considered{, though the efficiency of feedback is also sensitive to the value of $N_{\rm FB}$ used due to numerical overcooling \citep{dallavecchia12}. In EAGLE, feedback energy is stored until gas can be heated by at least some fixed temperature \citep{booth09}.}
The effective $\alpha$ that enters equation \eqref{bondi} is more varied, and is not constant in EAGLE.
However, we showed in \citet{taylor14} that the value of $\alpha$ does not affect the growth of black holes since the feedback is self-regulating, and so the different values of $\alpha$ in Table \ref{tab:agn} should not influence our conclusions.
Finally, we note that Illustris and IllustrisTNG include non-thermal feedback modes.
Both adopt a kinetic mode, and Illustris has a further radiative feedback mode whereby the cooling rate of the gas is changed by changing its ionisation state.
It is much less clear how these differences in the feedback model compared to TK affect the growth of black holes, however, since the models are calibrated to produce realistic galaxies we expect the impact on black hole growth to be small.}

\subsection{Data analysis}
Given the different seeding conditions between these simulations, to make a comparison, we record $\mbh$ when $\mdm = \dmn$ in TK-IC1 and TK-IC2 and compare the mean and distribution of $\mbh$ to the seed mass of other simulations. We define $\mbhdm$ as $\mbh$ when $\mdm = \dmn$, $\mdmbh$ as $\mdm$ when $\mbh = \bhn$, $\zdm$ as the redshift when $\mdm = \dmn$, and $\zbh$ as the redshift when $\mbh = \bhn$.

The gas, stellar, and black hole particles are associated with the group of their nearest dark matter particle neighbour. These groups are identified as galaxies in TK-IC1 and TK-IC2. However, we group dark matter particles within 0.02 times the mean seperation distance between dark matter particles together, this small seperation distance does not allow for an accurate determination of dark matter halo mass. Therefore, we adopt $M_{200}$, defined as the mass within a spherical region centred on the galaxy whose average density is 200 times the critical density of the Universe, as the dark matter halo mass for each galaxy.

To make a comparison between the seeding method of the different simulations, we look at $\mbhdm$. In Illustris, $\mbh = \bhn$ when $\mdm = 5 \times \dmn$; in IllustrisTNG $\mbh = 8 \times \bhn$ when $\mdm = 5 \times \dmn$; and in EAGLE, $\mbh = \bhn$ when $\mdm = \dmn$ (Table ~\ref{tab:seeding}). In TK-IC1 and TK-IC2, we extract the black hole mass once the dark matter halo has grown to $\dmn$ to compare to EAGLE, and $5 \times \dmn$ to compare to Illustris and IllustrisTNG. \footnote{The code used to analyse the data files is available on GitHub at: \url{https://github.com/ellawang44/bh_seeding}. Data files are available upon request.}

\section{Results}
\label{res}
\begin{figure*}
  \centering
  \subcaptionbox{}{\includegraphics[width=3.45in]{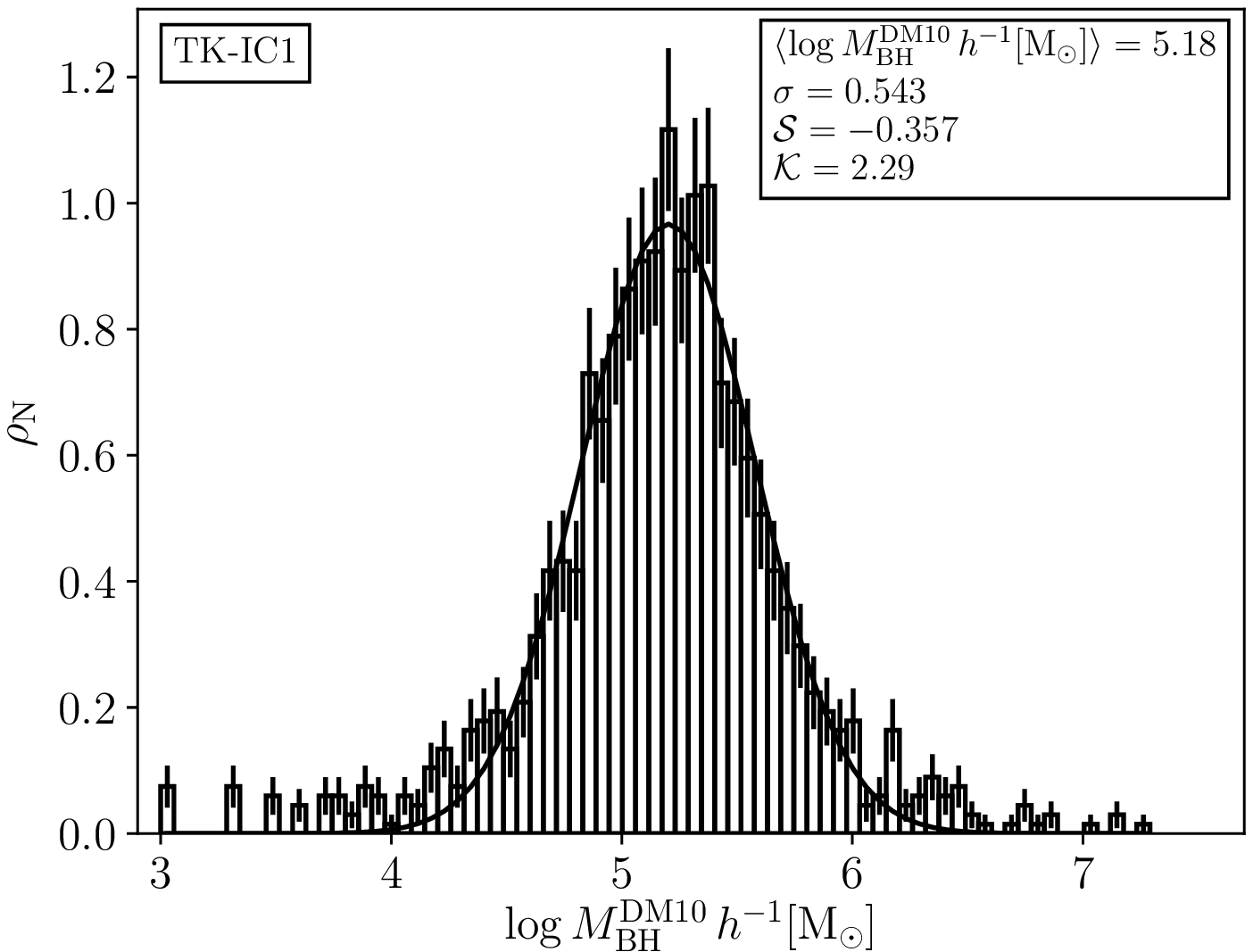}}\hspace{0em}%
  \subcaptionbox{}{\includegraphics[width=3.45in]{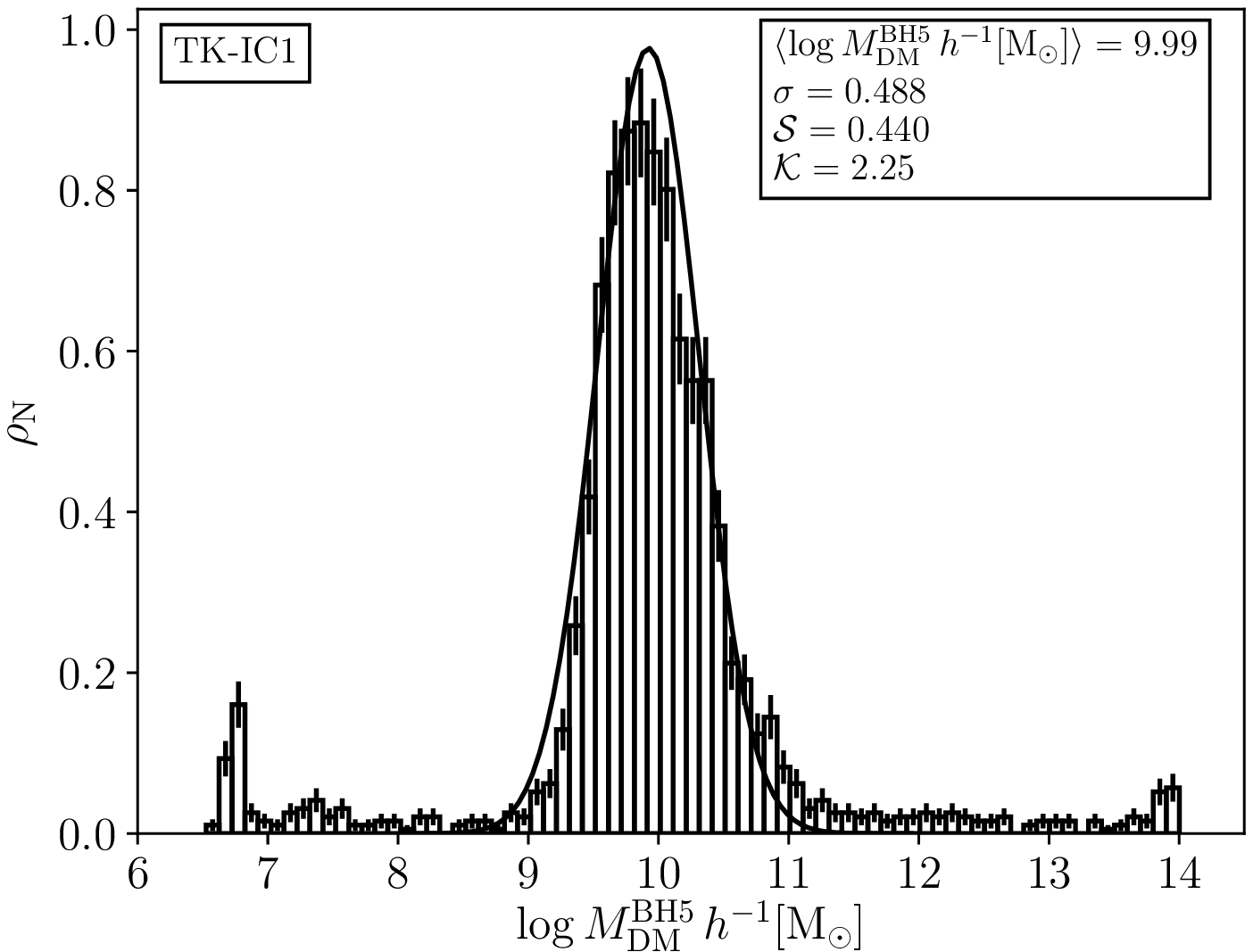}}
  \vspace{0em}%
  \subcaptionbox{}{\includegraphics[width=3.45in]{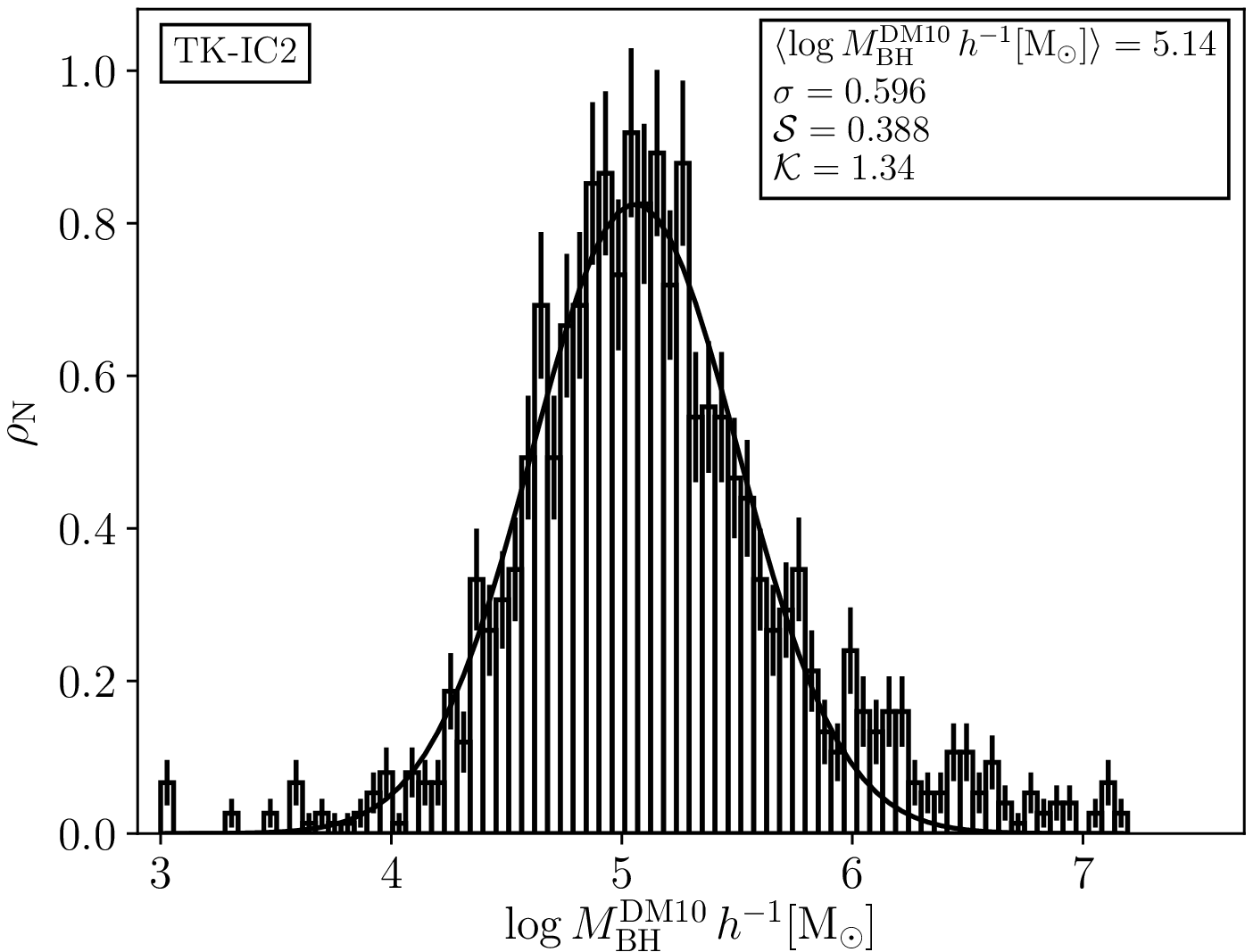}}\hspace{0em}%
  \subcaptionbox{}{\includegraphics[width=3.45in]{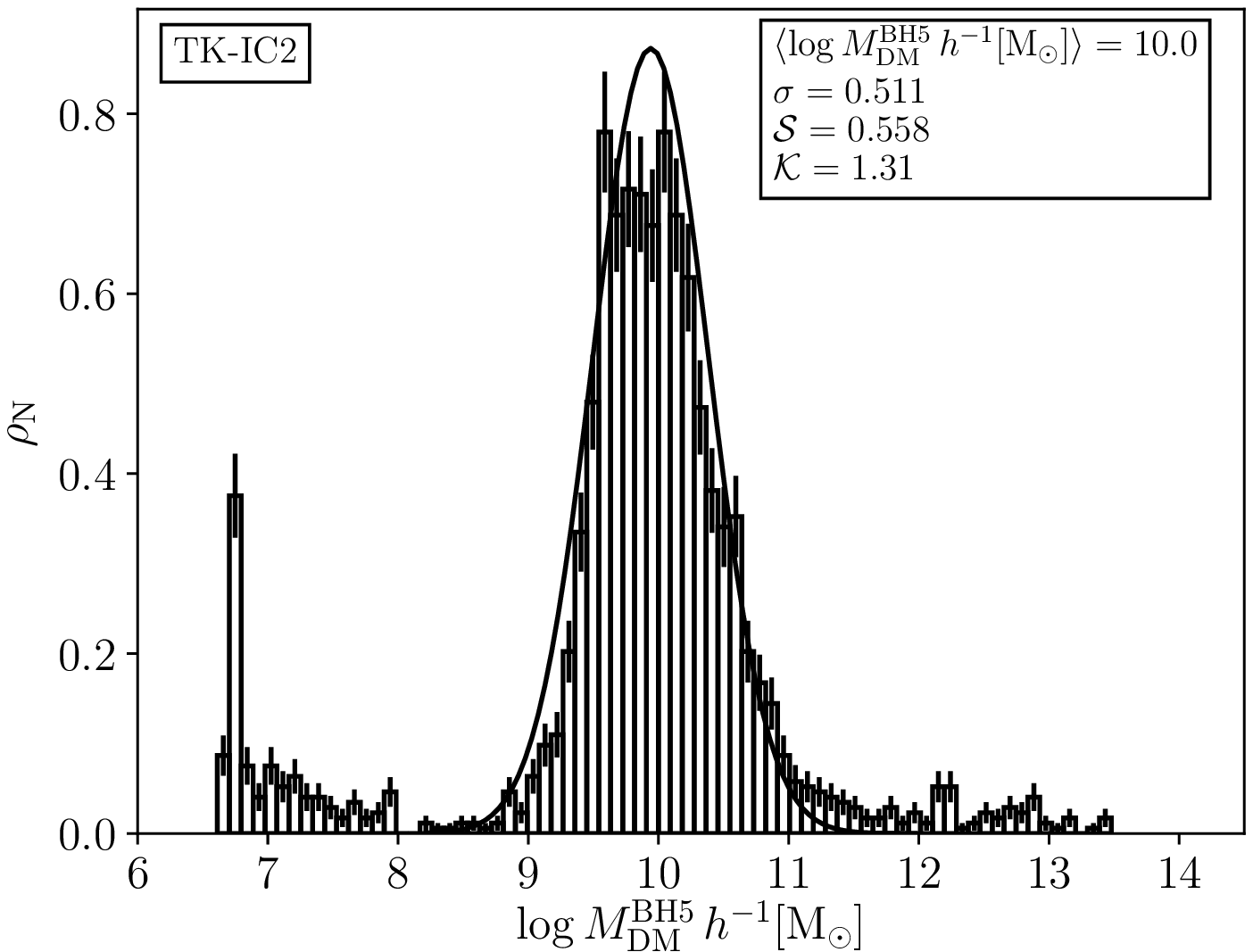}}
  \caption{The distribution of $\mdmbh$ (TK-IC1 in top-left panel, TK-IC2 in bottom-left panel) and $\mbhdm$ (TK-IC1 in top-right panel, and TK-IC2 in bottom-right panel). The solid line is a 2nd order Gaussian distribution (equation \ref{gauss_2nd_order}) fit the data set (the fitted values are reported in Table \ref{tab:data_stats}). The vertical lines on top of each bin are the Poisson errors given by $\sqrt{N}$ scaled to a number density, where $N$ is the number of galaxies in the bin.}
    \label{fig:key_findings}
\end{figure*}

We are interested in the correlation between black hole growth and dark matter halo growth, particularly in the mass region where black holes are typically seeded (see Table~\ref{tab:seeding} for a comparison). {The probability density distribution of} $\mbhdm$ in TK-IC1 is shown in the top-left panel, and {the distribution of} $\mdmbh$ is shown in the top-right panel of Fig.~\ref{fig:key_findings}; $\mbhdm$ in TK-IC2 is shown in the bottom-left panel, and $\mdmbh$ is shown in the bottom-right panel of Fig.~\ref{fig:key_findings}. A 2nd order Gaussian distribution given by
\begin{equation}
f(x) = \exp(a_0 + a_1 x + a_2 x^2)
\label{gauss_2nd_order}
\end{equation}
is fit to the data in Fig.~\ref{fig:key_findings}. The fit is shown with a solid line, fitted values are reported in Table \ref{tab:data_stats}. In the top-right and bottom-right panels of Fig.~\ref{fig:key_findings}, we do not fit galaxies with
$\mdm < 10^8 \, h^{-1} \, \msun$ or $\mdm > 10^{12} \, h^{-1} \, \msun$. This is because galaxies with low $\mdm$ are near the resolution limit of the simulation, {galaxies with $\mdm > 10^{12}\, \msun$ are almost all satellite galaxies}. The mean, standard deviation, skewness, and kurtosis (where kurtosis = 0 and skewness = 0 for a Gaussian distribution) of the data set is shown in Fig.~\ref{fig:key_findings} and reported in Table \ref{tab:data_stats}.

\begin{figure*}
  \centering
  \subcaptionbox{}{\includegraphics[width=3.45in]{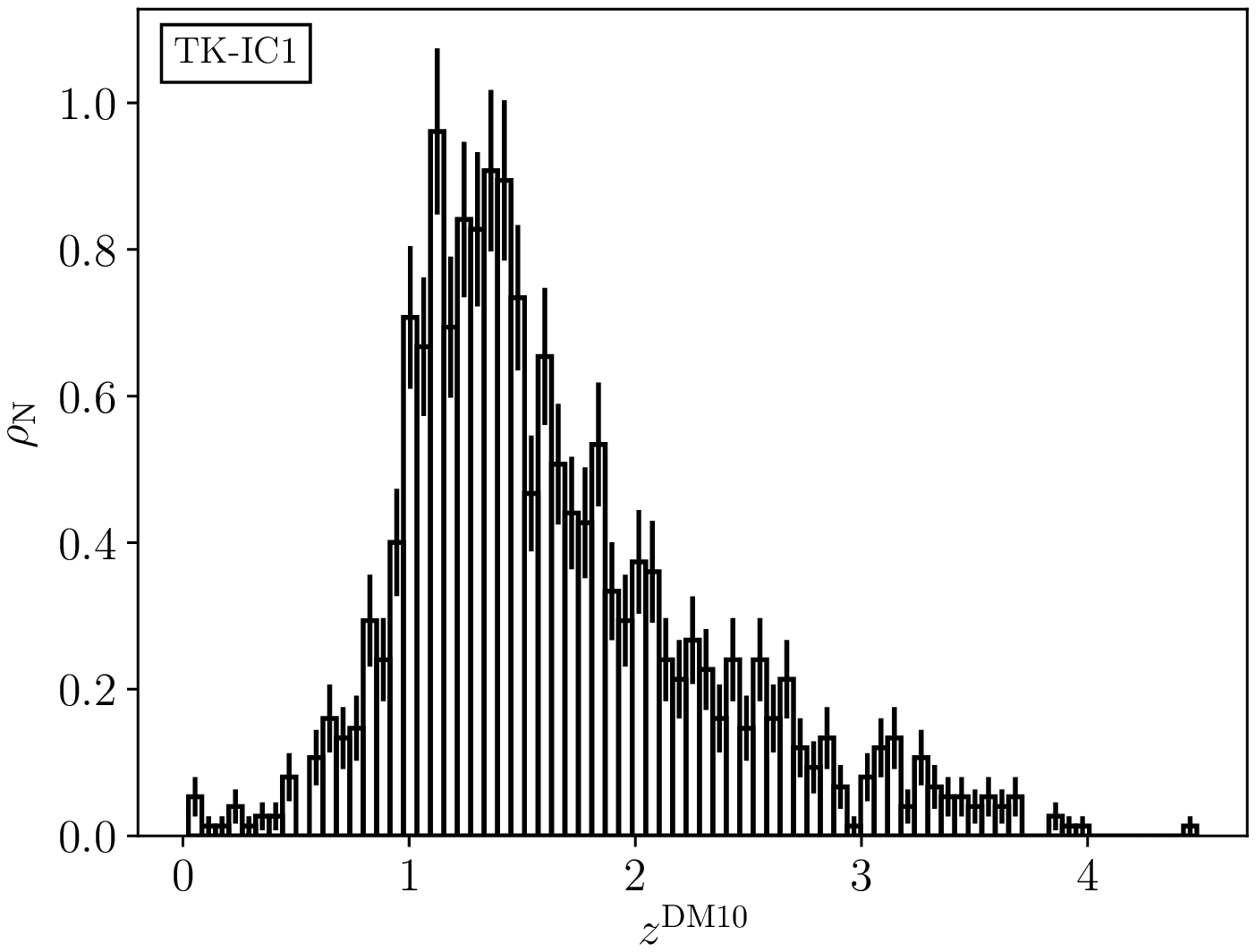}}\hspace{0em}%
  \subcaptionbox{}{\includegraphics[width=3.45in]{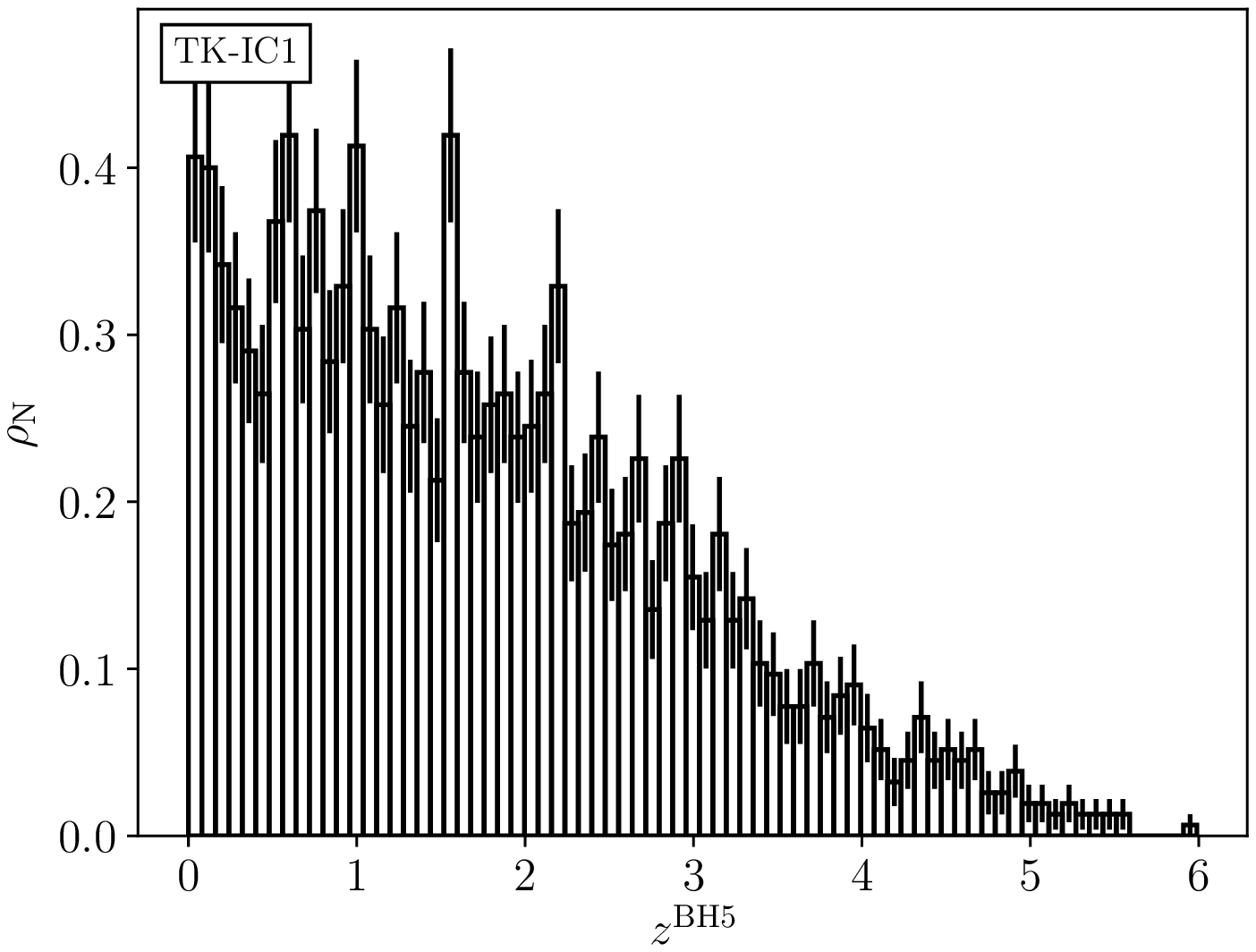}}
  \caption{The redshift distributions $\zdm$ (left-hand panel) and $\zbh$ (right-hand panel). The vertical lines on top of each bin are the Poisson errors given by $\sqrt{N}$ scaled to a number density, where $N$ is the number of galaxies in the bin. This data is from TK-IC1 with black holes of seed mass $\sm$, using gas properties as the seeding condition.}
	\label{fig:z}
\end{figure*}

\begin{table*}
\setlength{\tabcolsep}{5.0pt}
\begin{tabular}{lcccc}
\hline
& \multicolumn{2}{c}{$\log \mbhdm$} \quad & \multicolumn{2}{c}{$\log \mdmbh$} \\
& TK-IC1 & TK-IC2 & TK-IC1 & TK-IC2 \\
\hline
Mean & $5.18$ & $5.14$ & $9.99$ & $10.0$ \\
Standard deviation & $0.543$ & $0.596$ & $0.488$ & $0.511$ \\
Skewness & $-0.357$ & $0.388$ & $0.440$ & $0.558$ \\
Kurtosis & $2.29$ & $1.34$ & $2.25$ & $1.31$ \\
$a_0$ & -94.7 & -64.0 & -321 & -250 \\
$a_1$ & 36.4 & 25.2 & 64.6 & 50.3 \\
$a_2$ & -3.50 & -2.49 & -3.26 & -2.53 \\
\hline
\end{tabular}
\caption{Mean, standard deviation, skewness, and kurtosis of the distributions shown in Fig.~\ref{fig:key_findings}. The fitted parameters of the 2nd order Gaussian given by equation \ref{gauss_2nd_order}.}
\label{tab:data_stats}
\end{table*}

We find that most black holes in TK-IC1 are consistent with those in other simulations. However, strictly speaking, most black holes in TK-IC1 are slightly more massive than in Illustris and EAGLE when the dark matter halo reaches $\dmn$, as shown in the top-left panel of Fig.~\ref{fig:key_findings}. Therefore, when the black holes in this simulation reach $\bhn$ the associated dark matter haloes should be smaller than $\dmn$, which is observed in the top-right panel of Fig.~\ref{fig:key_findings}. These results imply that black holes reach $\bhn$ earlier than dark matter haloes reach $\dmn$. This is reflected in Fig.~\ref{fig:z}, which shows $\zdm$ in the left panel and $\zbh$ in the right panel. Comparatively, more dark matter haloes reach $\dmn$ than black holes reach $\bhn$ at lower redshifts, indicating that dark matter haloes in TK-IC1 are smaller than those in other simulations for $\mbh = \bhn$.

In order to test whether these results are due to cosmic variance or not, we ran the simulation again with different initial conditions and repeated the analysis (TK-IC2 in Table~\ref{tab:seeding}). The distributions are shown in the bottom-left and bottom-right panels of Fig.~\ref{fig:key_findings}, with the first four moments of these data sets reported in Table~\ref{tab:data_stats}. We find that the mean and standard deviation of the histograms depend very little on the initial conditions, both values differing by $\leq 0.1$. The skewness and kurtosis values are dependent on initial conditions, with skewness varying by $\sim 1$ and kurtosis varying by $>1$ between the simulations.

\section{Discussion}
\label{diss}
\subsection{Differences due to seeding methods}
In the previous section, we showed that $\mbhdm$ is consistent ($\langle \log \mbhdm \rangle = 5.18$) even though we apply a different seeding method compared to other simulations. {However, the evolutionary histories of individual galaxies could still be different, and we show the differences in this subsection. Since the detailed properties of galaxies necessary for our analysis are not available for EAGLE or Illustris simulations, we show the impact of seeding with our simulations, keeping the initial conditions and other sub-grid physics exactly the same. Any difference we show here is directly or indirectly caused by the difference in the seeding.}

In the Bondi-Hoyle accretion model, $\dot{M}_{\textrm{acc}} \propto \mbh^2$ (equation~\ref{bondi}).
Therefore, assuming that the properties of the gas local to the black holes are the same in the different simulations, the factor by which the accretion rates differ is
\begin{equation}
\left( \frac{ \langle \mbhdm \rangle}{\bhn} \right)^2 = 2.25.
\end{equation}
In Illustris and IllustrisTNG, black holes are seeded when $\mdm =5\times \dmn$; in TK-IC1 this corresponds to $\langle M_{\rm BH} \rangle=10^{5.26}\,h^{-1}\, \msun$, and our estimate for the ratio of the accretion rates is 3.24 and 0.0506, respectively.
These data are given in Table \ref{tab:accretion}.

\begin{table*}
\setlength{\tabcolsep}{5.0pt}
\begin{tabular}{lccc}
\hline
& TK-IC1 & Seed Mass & Factor difference in accretion rate \\
\hline
$\mdm = 1 \times \dmn$ (EAGLE) & $\langle \mbh \rangle = 10^{5.18}\, h^{-1}\, \msun$ & $\mbh = \bhn$ & 2.25 \\
$\mdm = 5 \times \dmn$ (Illustris) & $\langle \mbh \rangle = 10^{5.26}\, h^{-1}\, \msun$ & $\mbh = \bhn$ & 3.24 \\
$\mdm = 5 \times \dmn$ (IllustrisTNG) & $\langle \mbh \rangle = 10^{5.26}\, h^{-1}\, \msun$ & $\mbh = 10^{5.90} \, h^{-1} \, \msun$ & 0.0506 \\
\hline
\end{tabular}
\caption{Difference factor of accretion rate of black holes under the same $\mdm$ in the various simulations. For near-seed mass black holes, there exists a difference in mass, and thus a difference in accretion rate, assuming that these black holes inhabit galaxies with similar environments. Black holes in the Illustris simulation will grow the fastest, however, they will also release the most feedback energy, and thus may have longer periods where they self-regulate.}
\label{tab:accretion}
\end{table*}

For EAGLE and Illustris, the accretion rate ratios are close to 1, suggesting that the details of the seeding method do not strongly affect the subsequent evolution of the host galaxy.
For IllustrisTNG, the ratio is much smaller, but this discrepancy may not last to late times since AGN feedback is self regulating \citep[e.g.,][]{sijacki07,dubois12,taylor14,schaye15,volonteri16}.
To test the effects of the seeding method on galaxy evolution more explicitly, we ran a simulation with the same initial conditions as TK-IC1, and a seeding mechanism that matched as closely as possible the one used in the Illustris and EAGLE simulations, leaving all other physics unchanged.
We have developed our own subroutine of a parallel FoF finder, which was run on-the-fly during the simulation, with a linking length of 0.2 to find haloes whose dark matter mass exceeded $\dmn$; the gas particle closest to the centre of mass of such groups was converted into a black hole with seed mass $10^5\,h^{-1}\msun$. This simulation was evolved to $z\sim1$, and the most massive galaxies were matched between this simulation and TK-IC1 for direct comparison. {This match was done in both position and mass.} We refer to this new simulation as TK-IC1-FoF.

\begin{figure}
	\centering
	\includegraphics[width=0.48\textwidth,keepaspectratio]{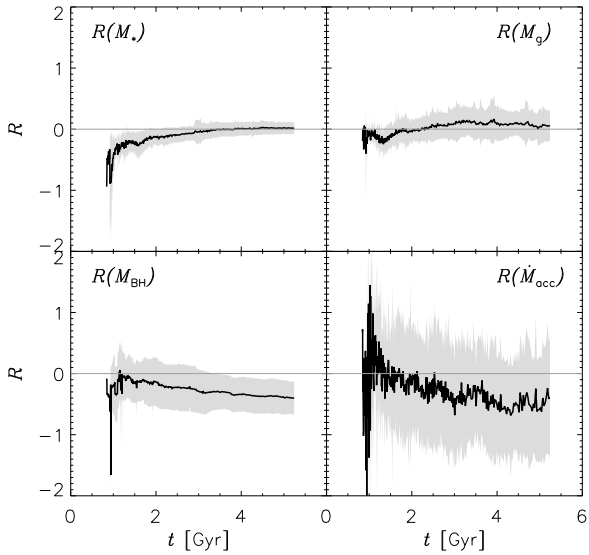}
	\caption{Evolution of $R$ (equation \ref{eq:ratio}) with cosmic time. From top left to lower right, the panels are for stellar mass ($M_*$), gas mass ($M_{\rm g}$), black hole mass ($M_{\rm BH}$), and black hole accretion rate ($\dot{M}_{\rm acc}$). The mean and $\pm 1\sigma$ are shown.}
	\label{fig:FOFCompare}
\end{figure}

We define the quantity
\begin{equation}\label{eq:ratio}
	R\left(X\right)=\log\left(X_{\rm TK-IC1}/X_{\rm TK-IC1-FoF}\right),
\end{equation}
for $X=M_*,\,M_{\rm g},\,M_{\rm BH},\,\dot{M}_{\rm acc}$ to compare the evolution of galaxy properties.
In Fig. \ref{fig:FOFCompare}, we show $R$ averaged across galaxies as a function of time (lines correspond to mean values and $\pm1\sigma$).
In the top-left panel of Fig. \ref{fig:FOFCompare}, we see that the stellar mass of galaxies is insensitive to the seeding method, especially at later times.
At early times, galaxies in TK-IC1 contain numerous, low-mass black holes that efficiently delay star formation compared to TK-IC1-FoF where only a single, central black hole is present.

\begin{figure}
	\centering
	\includegraphics[width=0.48\textwidth,keepaspectratio]{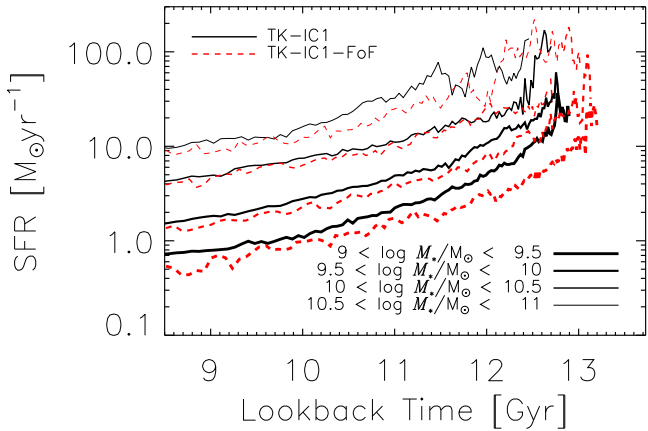}
	\caption{{Median SFR as a function of time for galaxies in different mass bins in the simulations TK-IC1 and TK-IC1-FoF (solid black and dashed red lines respectively).}}
	\label{fig:sfr_delay}
\end{figure}

{Fig. \ref{fig:sfr_delay} shows the median SFR of galaxies in a range of mass bins\footnote{Galaxies are binned by their stellar mass at a given redshift, and can be present in more than one mass bin at different times.} for the simulations TK-IC1 and TK-IC1-FoF (solid black and dashed red lines, respectively). The delay in star formation is most apparent in the lowest mass bin; as galaxies grow in TK-IC1, black holes merge, and most massive galaxies host only one black hole \citep{taylor14}. In the high mass bin, different strengths of AGN feedback due to the different black hole masses (see below) cause the small difference between the two simulations.

Analysing the other panels of Fig.~\ref{fig:FOFCompare},} $M_{\rm g}$ is also, on average, not strongly affected by seeding mechanism (top right panel).
However, $M_{\rm BH}$ is not consistent between the two simulations, with the discrepancy growing with time (lower left panel), and the galaxies of TK-IC1 hosting higher mass black holes at higher redshifts and lower mass black holes at $z \sim 1$.
The same trend is seen also for black hole accretion rate (lower right panel), though with much larger scatter.

These results seem to be at odds with Fig.~\ref{fig:key_findings}, in which $\mbhdm$ is slightly greater for TK-IC1 than a simulation with a FoF-based black hole seeding scheme. However, Fig.~\ref{fig:key_findings} shows the $\mbh$ for the same $\mdm$ at different times, whilst Fig.~\ref{fig:FOFCompare} shows the $\mbh$ for the same times but different $\mdm$. Fig.~\ref{fig:FOFCompare} shows that the subsequent growth of the black holes is faster in TK-IC1-FOF than TK-IC1. Our earlier assumption that the gas properties local to the black hole in different simulations are the same is overly simplistic, and it is likely that early feedback from low-mass black holes in TK-IC1 lowers the gas density in galaxy centres (but does not expel gas from the galaxy; see top right panel of Fig.~\ref{fig:FOFCompare}), causing a lower accretion rate. With the FoF-based seeding method, the gas is able to build up in the galaxy centre before the black hole forms, leading to higher accretion rates and faster initial black hole growth. Therefore, the black holes, and by extension, the galaxies, evolve differently given different seeding methods due to the different conditions of the galaxies when the black holes are seeded.

\begin{figure}
	\centering
	\includegraphics[width=0.48\textwidth,keepaspectratio]{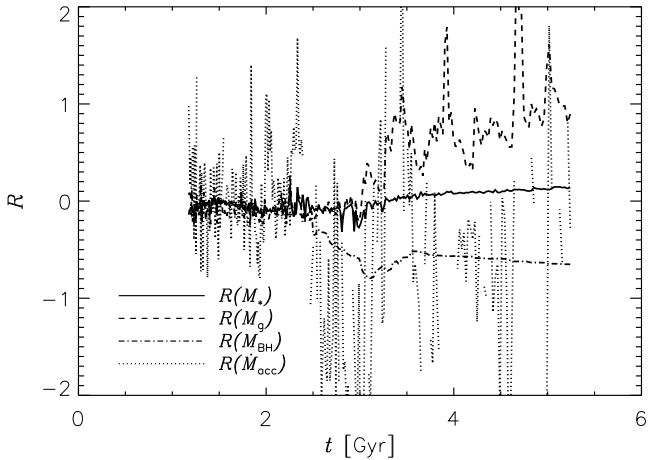}
	\caption{Evolution of $R$ (equation \ref{eq:ratio}) with cosmic time for a single galaxy that experiences a major merger at $t=2.5$ Gyr. The solid line is for $M_*$, dashed for $M_{\rm g}$, dot-dashed for $M_{\rm BH}$, and dotted for $\dot{M}_{\rm acc}$.}
	\label{fig:FOFCompare0}
\end{figure}

Galaxy mergers are highly non-linear, and can exacerbate and exaggerate the differences in black hole growth in simulations with different seeding methods. We illustrate this in Fig. \ref{fig:FOFCompare0}, which shows $R$ as a function of time for a single galaxy.
At $t<2.5$ Gyr, $R\approx0$ for all properties, meaning that the galaxy is evolving similarly in the separate simulations.
The galaxy experiences a major merger at $t=2.5$ Gyr, after which the black hole masses diverge in the different simulations.
At $t\approx3$ Gyr, feedback from the rapid growth of the black hole in TK-IC1-FoF begins to affect the total gas content of the galaxy, and, to a smaller extent, its stellar mass.
The expulsion of gas due to AGN feedback following the merger may have profound effects on the subsequent chemical evolution of both the galaxy and the intergalactic medium \citep[IGM;][]{taylor15b}; this will be investigated in detail in a future work.

\subsection{The origin of downsizing}
In the FoF seeding, black holes are seeded at a higher redshift in galaxies with a bigger $\mdm$ compared to galaxies with a smaller $\mdm$, and thus cause AGN feedback earlier. However, with our seeding method, we do not choose the galaxy halo mass and thus there is no guarantee that we have downsizing due to quenching. In this subsection, we show that we do have downsizing with our method and explain the origin.

\begin{figure*}
    \centering
		\begin{subfigure}[t]{0.33\textwidth}
			\includegraphics[width = \linewidth]{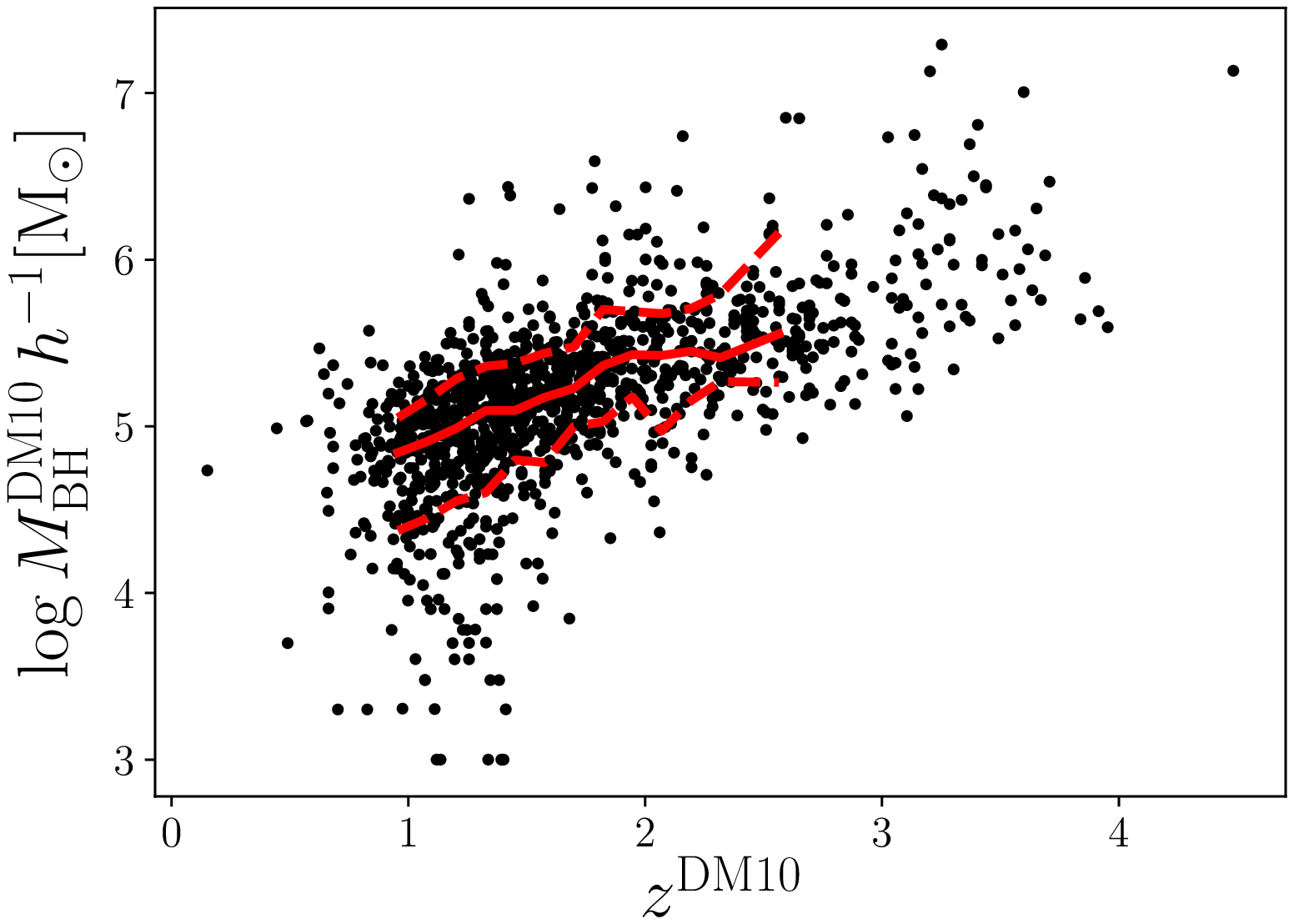}
		\end{subfigure}
		\hfill
   	    \begin{subfigure}[t]{0.33\textwidth}
			\includegraphics[width = \linewidth]{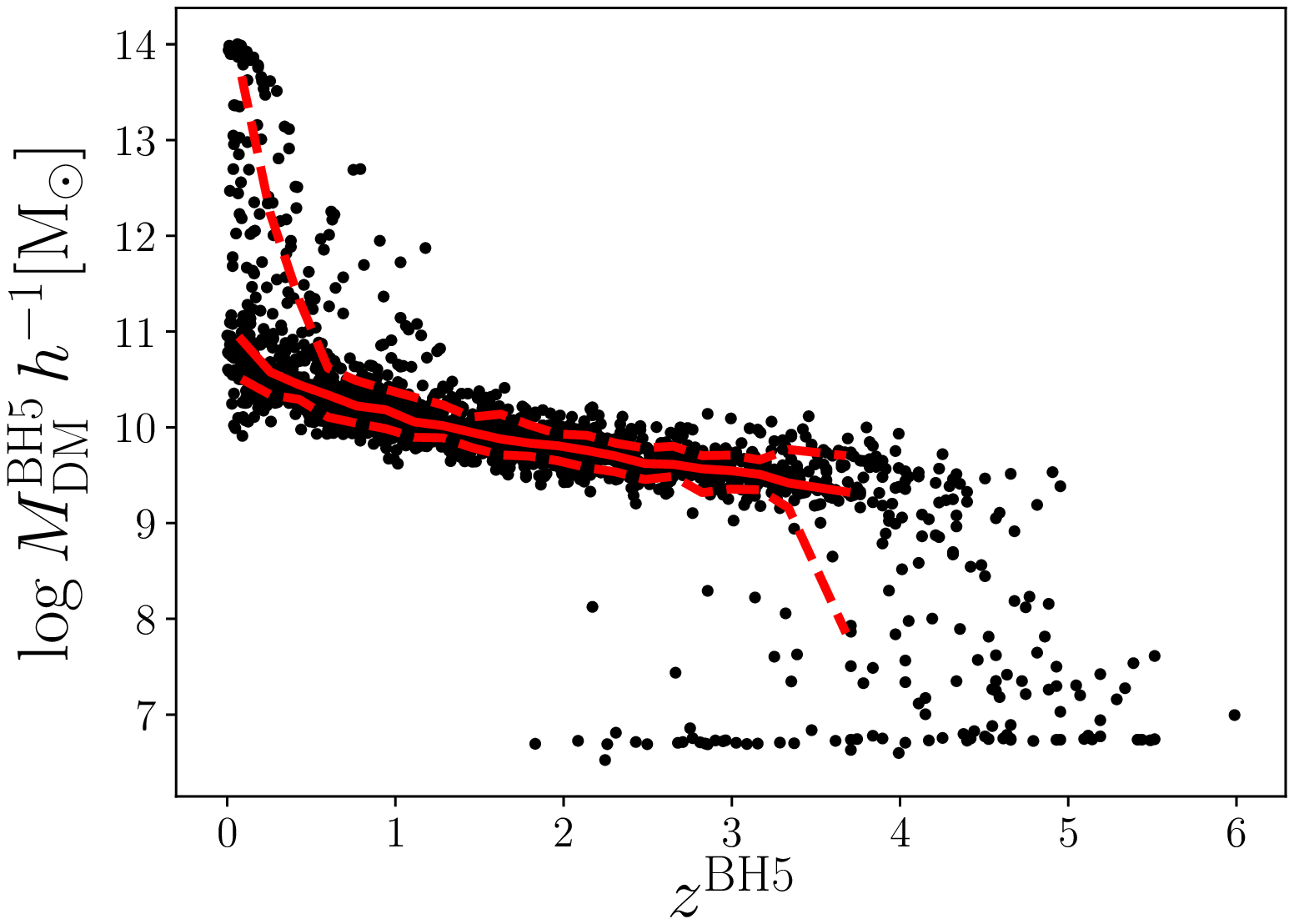}
		\end{subfigure}
		\hfill
		\begin{subfigure}[t]{0.33\textwidth}
			\includegraphics[width = \linewidth]{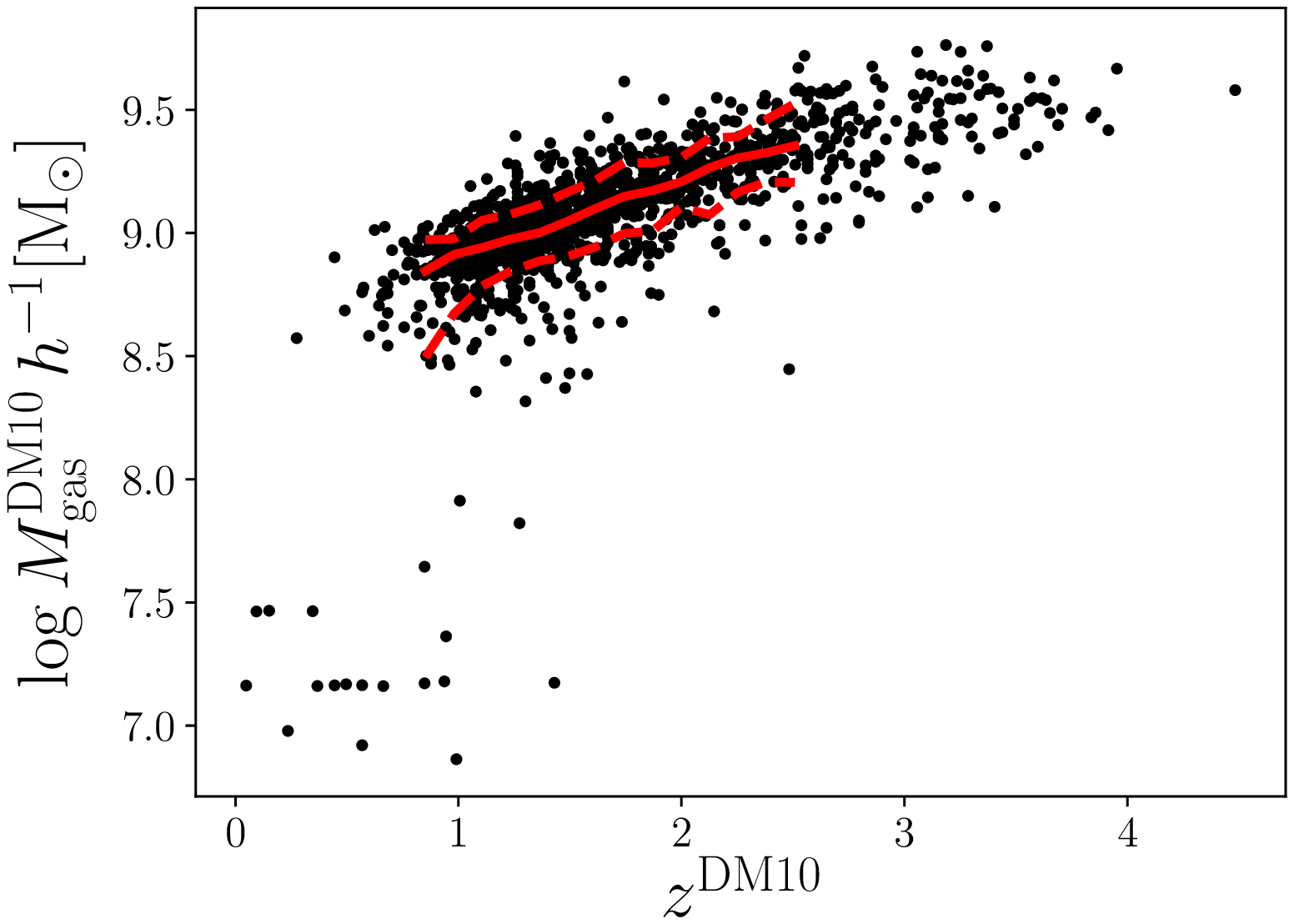}
		\end{subfigure}
		\caption{$\mbhdm$ of galaxies as a function of $\zdm$ is shown in the left panel. $\mdmbh$ of galaxies as a function of $\zbh$ is shown in the center panel. $\mgdm$ of galaxies as a function of $\zdm$ is shown in the right panel. Different properties of the same galaxies are shown in the left and right panel. The median of the points is given by the solid line, whilst the dotted lines show the 16th and 84th percentiles. The median, 16th, and 84th percentiles are calculated by splitting the redshift into 35 equally spaced bins. We only calculate the median, 16th, and 84th percentiles if there are more than 30 data points in the bin.}
		\label{fig:ds}
\end{figure*}

For TK-IC1, we consider the relationship between $\mbhdm$ and $\zdm$, shown in the left panel; and the relationship between $\mdmbh$ and $\zbh$, shown in the center panel of Fig.~\ref{fig:ds}. These figures reflect the redshift evolution of the same galaxies shown in the top-left panel of Fig.~\ref{fig:key_findings} and the redshift evolution of the top-right panel of Fig.~\ref{fig:key_findings} respectively. The upturn at lower redshifts in the center panel of Fig.~\ref{fig:ds} is due to satellite galaxies: the dark matter haloes of satellite galaxies are not separately identified from their host galaxies. The downturn at higher redshifts in the same figure is due to the resolution limit of the simulation. The left panel of Fig.~\ref{fig:ds} shows that $\mbhdm$ is larger for larger $\zdm$. Thus, we see that black holes grow more quickly compared to their dark matter haloes at higher redshifts. The middle panel of Fig.~\ref{fig:ds} echoes this result, showing that galaxies with larger $\zbh$ have lower $\mdmbh$.

The right panel of Fig.~\ref{fig:ds} shows the relationship between $\mgasdm$ and $\zdm$ of the same galaxies as those shown in the left panel of Fig.~\ref{fig:ds}, where we observe the same increasing trend: higher $\zdm$ galaxies contain more gas mass. Also at higher $\zdm$, the physical separation of particles is smaller (due to the expansion of the universe). This means that the density, $\rho$, of gas is higher in high-redshift galaxies.
The Bondi-Hoyle accretion rate gives us that $\dot{M}_{\rm{acc}} \propto \rho \mbh^2$ (equation.~\ref{bondi}). Both $\rho$ and $\mbhdm$ are larger for high-redshift galaxies, implying that the accretion rates of these galaxies are higher. Thus, black holes in high-redshift galaxies grow faster than black holes in low-redshift galaxies. Since black holes in high-redshift galaxies have a larger $\mbhdm$, they quench star formation on a galactic scale before low-redshift black holes, leading to the observed downsizing effect.

\section{Summary and Conclusions}
State-of-the-art cosmological simulations seed black holes depending on dark matter halo mass or gas properties. These different seeding criteria produce black holes with different growth histories, and as a result, different accretion rates. The seed mass and thus accretion history of the black holes has significant impact on AGN feedback, especially in the early evolution of the Universe and the black hole's host galaxies.

In this study, we compared the predicted black hole masses from \citet{taylor14} to those imposed in the Illustris, IllustrisTNG, and EAGLE simulation suites, and discussed the implications of these results. We find that the black hole masses are consistent but the average is slightly larger in \citet{taylor14} than in the other simulations for galaxies with similar dark matter halo masses. Different seeding methods will produce populations of galaxies with statistically similar properties (such as $M_*$, $M_{\rm g}$) by $z=0$, but the evolution of individual galaxies can be very different due to mergers. These differences may manifest themselves in the chemical properties of both galaxies and the IGM, observations of which must be reproduced by simulations.

We also find evidence for downsizing in TK-IC1, whereby black holes in galaxies at higher redshifts are more massive and grow more quickly compared to simulations with black hole seeding based on dark matter halo mass. This means that with gas-based seeding of stellar-mass black holes (the seeding method used in TK-IC1), star formation is quenched earlier in high-redshift galaxies compared to low-redshift galaxies.

The black hole seeding criteria should be reconsidered in hydrodynamical simulations. In this study, we find the distribution functions of $\mbhdm$ for a physically motivated seeding model. These distribution functions could be implemented in future cosmological simulations by drawing seed black hole masses from the distributions provided here.

\section*{Acknowledgments}
{We thank the anonymous referee for their many useful suggestions that improved the quality of this paper.} C.F.~gratefully acknowledges funding provided by the Australian Research Council's Discovery Projects (grants DP150104329, DP170100603, and FT180100495) and the Australia-Germany Joint Research Cooperation Scheme (UA/DAAD). We thank for high-performance computing resources provided by the Leibniz Rechenzentrum and the Gauss Centre for Supercomputing (grants~pr32lo, pr48pi and GCS Large-scale project~10391), the Partnership for Advanced Computing in Europe (PRACE grant pr89mu), the Australian National Computational Infrastructure (grant~ek9), and the Pawsey Supercomputing Centre with funding from the Australian Government and the Government of Western Australia, in the framework of the National Computational Merit Allocation Scheme and the ANU Allocation Scheme. This work used the DiRAC Data Centric system at Durham University, operated by the Institute for Computational Cosmology on behalf of the STFC DiRAC HPC Facility (www.dirac.ac.uk). This equipment was funded by a BIS National E-infrastructure capital grant ST/K00042X/1, STFC capital grant ST/K00087X/1, DiRAC Operations grant ST/K003267/1 and Durham University. DiRAC is part of the National E-Infrastructure.

\bibliography{ref}
\bibliographystyle{mn2e}

\end{document}